\documentclass[11pt,a4paper]{article}

\usepackage[margin=1in]{geometry}
\usepackage{amsmath,amssymb,mathtools}
\usepackage{amsthm}
\usepackage{xcolor}
\usepackage{tikz}
\usetikzlibrary{arrows.meta}
\usepackage{microtype}
\usepackage{hyperref}

\hypersetup{colorlinks=true,linkcolor=blue,citecolor=blue,urlcolor=blue}

\newtheorem{proposition}{Proposition}

\newcommand{\Cl}{\mathrm{Cl}}
\newcommand{\R}{\mathbb{R}}
\newcommand{\C}{\mathbb{C}}
\newcommand{\Z}{\mathbb{Z}}
\newcommand{\graden}[2]{\langle #1 \rangle_{#2}}
\newcommand{\rev}[1]{\widetilde{#1}}
\newcommand{\bs}[1]{\boldsymbol{#1}}
\newcommand{\Biv}{\mathcal{B}}
\newcommand{\dd}{\mathop{}\!\mathrm{d}}

\title{\textbf{A Geometric Journey Through the Aharonov--Bohm Effect:\\
Tonomura's Toroid, Pseudoscalar Phase, and Quantum Holonomy in Clifford Algebra}}

\author{
Andrew Powell\thanks{Email:
\href{mailto:andrew.powell@physics.oxford.ac.uk}
{\texttt{andrew.powell@physics.oxford.ac.uk}}}\\
Academic Visitor, Department of Physics, University of Oxford
}\date{November 5, 2026}

\begin{document}
\maketitle

\begin{abstract}
The magnetic Aharonov--Bohm (AB) effect is often introduced through the idealised infinite solenoid and a complex scalar phase. We give a geometric-algebra formulation based on the compact toroidal magnets shielded by a superconductor employed in the electron-holography experiments of Tonomura and collaborators. Working in the real Euclidean Clifford algebra $\Cl(3,0)$, we distinguish the unit volume pseudoscalar $I$ (grade 3), coordinate angles (scalars, grade 0), tangent vectors (grade 1), and directed surface bivectors (grade 2). The magnetic field is represented by the dual bivector $\Biv=I\bs B$, while the electromagnetic $U(1)$ phase is represented by the central unit multivector $e^{I\alpha}$ generated by the pseudoscalar $I$, rather than by a spatial rotor with a bivector generator. We establish the linked toroidal geometry, formulate magnetic flux and Stokes' theorem intrinsically in geometric algebra, and derive the holonomy element $U_\Gamma=\exp(Iq\Phi_B/\hbar)$ from $U(1)$ gauge covariance, with every sign fixed explicitly. We then express superconducting fluxoid quantization in
the same notation and obtain the binary electron holonomy
$U_\Gamma = (-1)^N$ in the ideal limit where the linked
magnetic flux equals $N h/(2e)$. Finally, a short conditional-state description of the
quantum source and field shows how a backreaction can
introduce an additional phase and reduce interference
visibility. The conditional-state overlap is represented
by a central scalar--pseudoscalar element; preservation
of the fluxoid sector alone does not guarantee unit coherence.
\end{abstract}


\section{Introduction}

Classical electromagnetism describes forces through
the local fields
$\bs B=\nabla\times\bs A$ and
$\bs E=-\nabla V-\partial_t\bs A$.
The magnetic Aharonov--Bohm (AB) effect shows that
quantum interference can also depend on magnetic
flux linked by paths along which the field vanishes.

In 1959, Aharonov and Bohm
\cite{AharonovBohm1959} showed that a confined
magnetic flux can produce an observable relative
phase between electron paths lying in a region
where the magnetic field vanishes:
\begin{equation}
\Delta\theta_{\text{AB}} = \frac{q}{\hbar}\oint_\Gamma \bs A \cdot \dd\bs x = \frac{q\Phi_B}{\hbar} \pmod{2\pi},
\label{eq:AB-formula}
\end{equation}
where $\Delta\theta_{\text{AB}}$ is the relative phase shift between the interferometric paths and $\Gamma$ is the closed loop enclosing magnetic flux $\Phi_B$ that represents those paths. The effect demonstrates that quantum interference can depend
on gauge-invariant global information even when the magnetic
field vanishes along the accessible electron paths. This
information is encoded in the closed-loop phase factor, or
holonomy, of the electromagnetic connection
\cite{WuYang1975,PeshkinTonomura1989}. It does not, by itself,
establish that the underlying electromagnetic interaction
is non-local.

The preceding discussion treats the vector potential as a
classical background. Section~\ref{sec:backreaction} considers
how a quantum description of the source and electromagnetic
field can introduce a backreaction and reduce interference
visibility.

\subsection{The Solenoid Debate and Tonomura's Toroid}
In Aharonov and Bohm's paper \cite{AharonovBohm1959}  and in standard textbooks \cite{Sakurai2017,Griffiths2018}, the AB effect is derived for an infinitely long straight solenoid. For an ordinary finite, open-ended solenoid,
magnetic return flux extends outside the winding
region, making leakage fields an important
experimental consideration.  The effect prompted both theoretical objections,
concerning gauge choices, phase single-valuedness
and boundary conditions, and experimental objections,
concerning leakage fields and electron penetration
into the magnetic material
\cite{BocchieriLoinger1978,Hiley2013,Osakabe1986}.

Electron-holography experiments in 1982 were followed by the superconductively shielded experiments of 1986 \cite{Tonomura1982,Tonomura1986,Osakabe1986,Tonomura2005}. Tonomura and collaborators fabricated micrometre-scale
toroidal ferromagnets. In the detailed 1986 study, the
magnetic rings had outer diameters of approximately
$5$--$5.5\,\mu\mathrm m$, while the surrounding
superconducting structures had outer diameters of
approximately $6$--$7\,\mu\mathrm m$
\cite{Osakabe1986}. In the definitive 1986 experiment, the toroidal magnetic core was completely covered by a superconducting niobium ($\mathrm{Nb}$) layer and an outer copper ($\mathrm{Cu}$) cladding. When cooled below $9.2\,\text{K}$:
\begin{enumerate}
    \item The superconducting niobium shield suppressed magnetic
flux leakage into the accessible exterior.  The leakage
flux just above the transition temperature was less than
\(h/(20e)\); below the transition it was estimated to be
far less than \(h/(200e)\), using the shield thickness and
London penetration depth \cite{Osakabe1986}.
We idealise this strongly suppressed exterior field as
\(\mathbf B=0\) in the calculations below.
    \item The Nb and Cu layers strongly suppressed coherent
electron penetration into the magnetic-field region.
The outer Cu coating also reduced charging and
contact-potential effects that could otherwise produce
electrostatic phase shifts \cite{Osakabe1986}.
\end{enumerate}
A coherent electron wave illuminated the specimen,
including both the central aperture and the exterior
of the toroid. An electron biprism overlapped the
object wave with a reference wave to record an
off-axis hologram, from which the relative phase
between the aperture and exterior regions was
determined \cite{Osakabe1986}.
The two-path construction below is an idealised
representation of this relative-phase comparison.

\subsection{Pedagogical Aims}
The intended readers are advanced undergraduate
and beginning graduate students familiar with
vector calculus, electromagnetic potentials and
two-path quantum interference. The Clifford-algebra
identities needed here are introduced explicitly,
and no prior knowledge of superconductivity is
assumed beyond the physical inputs stated in
Section~\ref{sec:superconductivity}.

The educational purpose is to connect the geometry
of a shielded toroidal magnet with local gauge
equivalence, closed-loop phase transport and the
observed interference contrast. In particular,
the treatment distinguishes flux confined within
the toroidal tube from flux through its aperture,
electromagnetic phase from spatial rotation, and
fluxoid quantisation from magnetic-flux quantisation.
The geometric-algebra notation keeps the relevant
orientations and multivector grades explicit.

Heras \cite{Heras2022} relates the AB phase for a
closed flux line to linking number using vector calculus. Our purpose
is complementary: to connect this geometry with
phase transport and superconducting fluxoid
quantisation in real Clifford algebra.

Standard derivations of the AB effect use complex
wavefunctions $\psi\sim e^{i\theta}$, with the complex
unit $i$ introduced algebraically. Here we represent that complex structure by the
central pseudoscalar of the real Euclidean Clifford
algebra $\Cl(3,0)$, while representing the magnetic
field by its dual bivector
\cite{Hestenes1966,Hestenes1984,HavelDoran2002,
Hestenes2003,DoranLasenby2003,Garling2011}.

The principal learning goals are to distinguish
magnetic surface geometry from electromagnetic
phase, explain how local gauge equivalence permits
nontrivial closed-loop transport, and to connect
superconducting fluxoid quantisation with the
ideal even--odd electron phase contrast.

\section{The Euclidean Clifford Algebra $\Cl(3,0)$}
\label{sec:Clifford}

Let $\{\bs\sigma_1,\bs\sigma_2,\bs\sigma_3\}$ be a right-handed orthonormal vector basis for Euclidean three-space. The associative geometric product of vectors is
\begin{equation}
\bs u\bs v=\bs u\cdot\bs v+\bs u\wedge\bs v,
\qquad
\bs\sigma_j\bs\sigma_k+\bs\sigma_k\bs\sigma_j=2\delta_{jk}.
\label{eq:geometric-product}
\end{equation}
The unit pseudoscalar is
\begin{equation}
I:=\bs\sigma_1\bs\sigma_2\bs\sigma_3,
\qquad I^2=-1,
\qquad \rev I = -I.
\label{eq:pseudoscalar}
\end{equation}

Every element of $\Cl(3,0)$ is a \emph{multivector} and has
a unique decomposition into scalar, vector,
bivector and pseudoscalar parts:
\[
M=a+\sum_{i=1}^{3}a_i\bs\sigma_i
+a_{12}\bs\sigma_1\bs\sigma_2
+a_{23}\bs\sigma_2\bs\sigma_3
+a_{31}\bs\sigma_3\bs\sigma_1
+a_{123}I.
\]
A grade-$r$ multivector has no components of
grades other than $r$.

A nonzero grade-$r$ multivector is called an
\emph{$r$-blade} if it can be expressed as a single exterior
product of $r$ vectors:
\[
A_r=v_1\wedge\cdots\wedge v_r.
\]
Scalars are regarded as $0$-blades.
In $\Cl(3,0)$ every nonzero homogeneous multivector
is a blade, although this is not true in arbitrary
dimension. 

\subsection{Reversion}

\emph{Reversion} $\rev{(\cdot)}$ is the anti-automorphism of $\Cl(3,0)$
that reverses the order of the factors in every geometric product:
\begin{equation}
\rev{(AB)}=\rev B\,\rev A,
\qquad
\rev{(a_1a_2\cdots a_r)}=a_r\cdots a_2a_1,
\label{eq:reversion-def}
\end{equation}
for all $A,B\in\Cl(3,0)$ and vectors $a_k$. A grade-$r$ multivector $A_r$ acquires only a sign under
reversion:
\begin{equation}
\rev{A_r}=(-1)^{r(r-1)/2}\,A_r,
\label{eq:reversion-blade}
\end{equation}
so scalars and vectors are fixed, $\rev 1=1$ and $\rev{\bs v}=\bs v$, while
bivectors and the pseudoscalar change sign,
\begin{equation}
\rev J=-J \ \text{ for every bivector } J,
\qquad
\rev I=-I,
\label{eq:reversion-signs}
\end{equation}
the latter being consistent with Eq.~\eqref{eq:pseudoscalar}. Three
consequences are used throughout this paper.

\emph{(i) Reversion is the geometric-algebra analogue of complex conjugation.}
On the central subalgebra $\operatorname{span}_\R\{1,I\}\cong\C$ it acts
exactly as conjugation:
\begin{equation}
\rev{\left(e^{I\alpha}\right)}
=\exp\!\left(\rev I\,\alpha\right)
=e^{-I\alpha}
=\left(e^{I\alpha}\right)^{-1},
\qquad \alpha\in\R,
\label{eq:reversion-conjugate}
\end{equation}
so every unit phase element $U$ satisfies $U\rev U=\rev U U=1$. 

For these central phase elements, reversion
represents the adjoint, so $U^{-1}=\rev U$.
This identity is not asserted for arbitrary
multivectors.

\emph{(ii) Reversion supplies a non-negative scalar norm.} Writing
$M=\sum_{r=0}^{3}M_r$ for the grade decomposition of a multivector, products
of blades of unequal grade contain no scalar part, and each blade is returned
to a non-negative scalar by reversion. Hence
\begin{equation}
\graden{M\rev M}{0}
=\sum_{r=0}^{3}M_r\rev{M_r}
=\lVert M_0\rVert^2+\lVert M_1\rVert^2+\lVert M_2\rVert^2+\lVert M_3\rVert^2
\;\geq\;0,
\label{eq:reversion-norm}
\end{equation}
with equality if and only if $M=0$: for a vector,
$M_1\rev{M_1}=\bs v^2$, and for a bivector $J=a\wedge b$,
$J\rev J=-J^2=\lVert a\wedge b\rVert^2$, while
$I\rev I=-I^2=1$.

For the common-spin wave multivectors used in
Section~\ref{sec:interference}, this scalar norm supplies
the usual modulus-squared probability rule.

\emph{(iii) Reversion is involutive:} $\rev{\rev M}=M$ for every multivector
$M$, as follows immediately from applying Eq.~\eqref{eq:reversion-def} twice.

\subsection{Properties of the Pseudoscalar}
In $\Cl(3,0)$, $I$ represents an oriented unit volume. Since the dimension is odd, $I$ commutes with every multivector, including vectors and bivectors. The centre is
$\operatorname{span}_{\R}\{1,I\}\cong\C$.
Unlike a plane-specific bivector, $I$ therefore
provides a central representation of the imaginary
unit. We use it for electromagnetic phase, while
bivectors represent oriented planes.

The pseudoscalar also implements vector--bivector duality:
\begin{equation}
I\bs\sigma_1=\bs\sigma_2\wedge\bs\sigma_3,
\qquad
I\bs\sigma_2=\bs\sigma_3\wedge\bs\sigma_1,
\qquad
I\bs\sigma_3=\bs\sigma_1\wedge\bs\sigma_2.
\label{eq:duality}
\end{equation}
For a magnetic field vector $\bs B$, define the magnetic bivector
\begin{equation}
\Biv:= I\bs B.
\label{eq:magnetic-bivector}
\end{equation}
The bivector $\Biv$ specifies the oriented plane dual to the magnetic field direction. This is the natural object for magnetic flux, which is evaluated on an oriented surface element.

\section{Cylindrical Coordinates: Scalars, Vectors, and Bivectors}

The position vector in cylindrical coordinates is
\begin{equation}
\bs x(\rho,\phi,z)=
\rho\cos\phi\,\bs\sigma_1+
\rho\sin\phi\,\bs\sigma_2+
z\,\bs\sigma_3.
\label{eq:cylindrical-position}
\end{equation}
The coordinate $\phi$ is a dimensionless scalar. The associated orthonormal vectors are
\begin{align}
\hat{\bs\rho}
&=\frac{\partial\bs x}{\partial\rho}
=\cos\phi\,\bs\sigma_1+\sin\phi\,\bs\sigma_2,\\
\frac{\partial\bs x}{\partial\phi}
&=\rho\hat{\bs\phi},\\
\hat{\bs\phi}
&=-\sin\phi\,\bs\sigma_1+\cos\phi\,\bs\sigma_2,\\
\hat{\bs z}
&=\frac{\partial\bs x}{\partial z}=\bs\sigma_3.
\end{align}
The displacement is the 1-vector
\begin{equation}
\dd\bs x=
\hat{\bs\rho}\,\dd\rho+
\rho\hat{\bs\phi}\,\dd\phi+
\hat{\bs z}\,\dd z.
\label{eq:line-element}
\end{equation}

Multiplication of the azimuthal unit vector by $I$ produces the directed meridional bivector plane:
\begin{align}
I\hat{\bs\phi}
&=-\sin\phi\,\bs\sigma_2\wedge\bs\sigma_3
+\cos\phi\,\bs\sigma_3\wedge\bs\sigma_1
\nonumber\\
&=\hat{\bs z}\wedge\hat{\bs\rho}.
\label{eq:azimuthal-duality}
\end{align}
Hence an azimuthal toroidal magnetic field $\bs B=B_\phi\hat{\bs\phi}$ has magnetic-bivector representation
\begin{equation}
\Biv=I\bs B=B_\phi\,\hat{\bs z}\wedge\hat{\bs\rho}.
\label{eq:toroidal-bivector}
\end{equation}

\section{Tonomura's Toroid and Gauss Linking Number}

A torus of major radius $R_0$ and minor radius $r_0<R_0$ is parametrised by
\begin{equation}
\bs x(\vartheta,\varphi)=
(R_0+r_0\cos\vartheta)\hat{\bs\rho}(\varphi)
+r_0\sin\vartheta\,\hat{\bs z},
\label{eq:torus}
\end{equation}
where $\varphi$ is the toroidal angle around the symmetry axis and $\vartheta$ is the poloidal angle around the tube cross-section. Its tangent vectors are
\begin{align}
\frac{\partial\bs x}{\partial\vartheta}
&=-r_0\sin\vartheta\,\hat{\bs\rho}
+r_0\cos\vartheta\,\hat{\bs z},\\
\frac{\partial\bs x}{\partial\varphi}
&=(R_0+r_0\cos\vartheta)\hat{\bs\phi}.
\end{align}
Their outer product yields the directed surface area bivector
\begin{equation}
\dd S_2=
\left(
\frac{\partial\bs x}{\partial\vartheta}
\wedge
\frac{\partial\bs x}{\partial\varphi}
\right)\dd\vartheta\,\dd\varphi.
\label{eq:torus-area-bivector}
\end{equation}
The parameter order $(\vartheta,\varphi)$ induces
the inward orientation of the torus surface.

\begin{figure}[t]
\centering
\begin{tikzpicture}[scale=1.1,>=Stealth]
  \node[above] at (1.0,3.0) {\textbf{Tonomura Toroidal Interferometer Geometry}};
  \draw[thick,fill=orange!20] (0,0) ellipse (3.2 and 1.25);
  \draw[thick,fill=white] (0,0) ellipse (1.35 and 0.52);
  \draw[dashed,blue!85!black,very thick,->] (0,0.08) ellipse (2.25 and 0.82);
  \draw[red!80!black,very thick,->] (-0.40,2.35) -- (-0.40,-2.35);
  \draw[green!55!black,very thick,->] (3.65,2.35) -- (3.65,-2.35);
  \draw[dashed,black!50] (-0.40,2.35) -- (3.65,2.35);
  \draw[dashed,black!50] (-0.40,-2.35) -- (3.65,-2.35);
  \draw[<->] (0,-0.65) -- (1.35,-0.65) node[midway,below] {\footnotesize $R_0-r_0$};
  \draw[<->] (0,-1.45) -- (3.2,-1.45) node[midway,below] {\footnotesize $R_0+r_0$};
  \node[draw,thick,fill=white,rounded corners,inner sep=6pt,align=left] at (6.3,0) {
    \textbf{Legend:}\\[3pt]
    {\color{blue!85!black}\textbf{---}} Toroidal Flux $\Phi_B$\\[3pt]
    {\color{red!80!black}\textbf{---}} Path $\gamma_1$ (Central Hole)\\[3pt]
    {\color{green!55!black}\textbf{---}} Path $\gamma_2$ (Outside Perimeter)\\[3pt]
    {\color{black!60}\textbf{- -}} Linked Loop $\Gamma=\gamma_1-\gamma_2$
  };
\end{tikzpicture}
\caption{Toroidal interferometer geometry. Electron branch $\gamma_1$
passes through the central aperture, while $\gamma_2$ passes outside
the outer perimeter. The trapped magnetic flux circulates azimuthally
within the toroidal ring (dashed blue curve) and does not thread the
aperture, which is field-free. The coloured paths show the central branch segments.
The dashed horizontal lines schematically indicate
the closures supplied by the omitted source and
detector connections; they are not additional electron
trajectories. For the once-linked configuration, the
completed difference contour
$\Gamma=\gamma_1-\gamma_2$ has linking number
$\ell=\pm1$ with the magnetic core. The drawing is a schematic projection: apparent
crossings do not indicate penetration of the
magnetic material.}
\label{fig:toroid}
\end{figure}
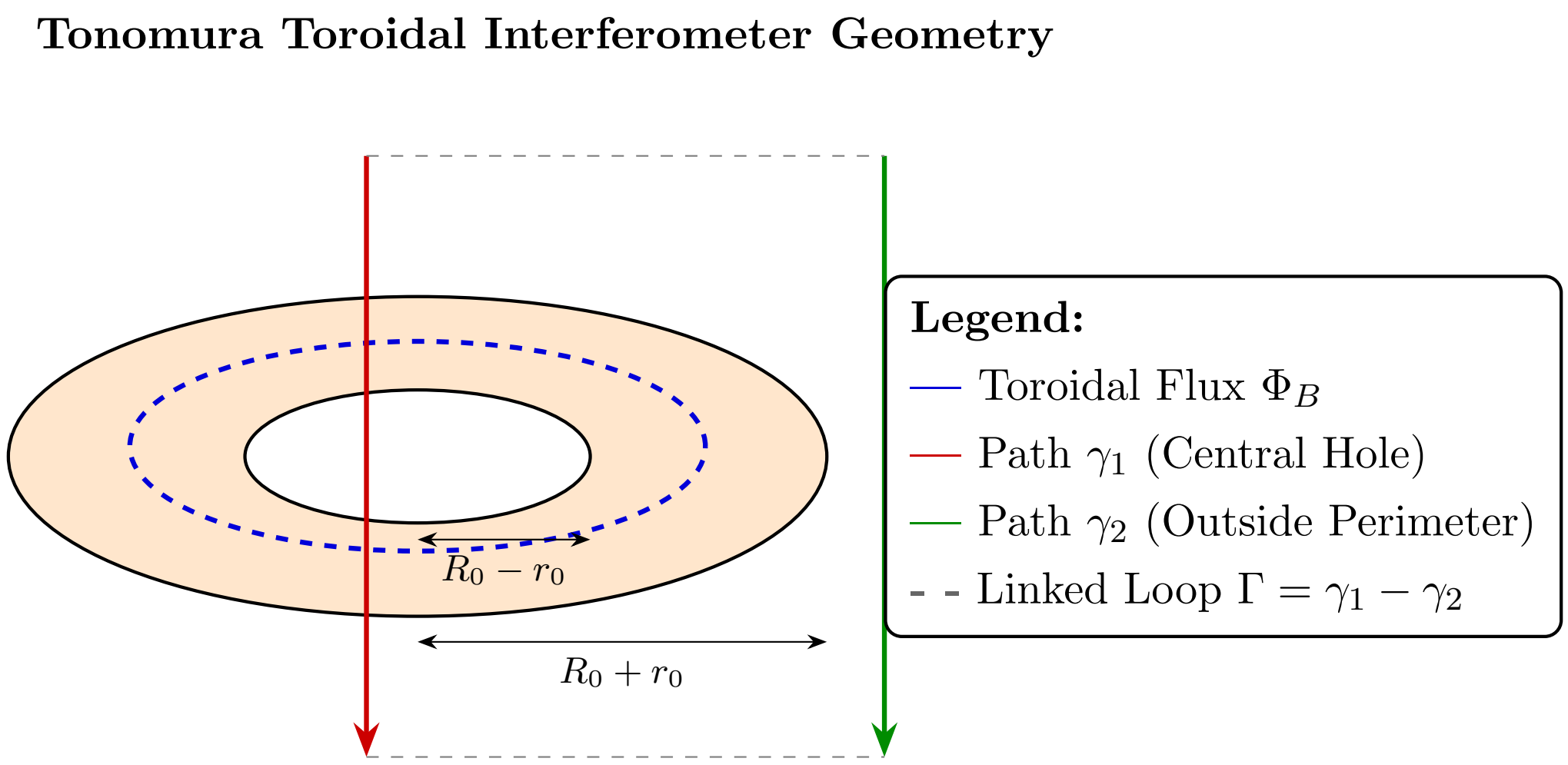

\subsection{Topological Beam Geometry}

Let $\gamma_1$ and $\gamma_2$ be complete paths from
a common source to a common detector. Their respective
central segments pass through the toroidal aperture
and outside the toroid. Figure~\ref{fig:toroid}
shows these central segments schematically; the
source and detector connections are suppressed.
The common endpoints allow the oriented closed
difference contour
\begin{equation}
\Gamma=\gamma_1-\gamma_2.
\label{eq:difference-contour}
\end{equation}
We consider the standard once-linked configuration,
with no additional winding of either branch and with
source and detector connections chosen as indicated
schematically in the figure. Let $C_{\rm mag}$ be a closed core curve following the toroidal magnetic flux. The Gauss linking number is
\begin{equation}
\ell(\Gamma,C_{\rm mag})
=
\frac{1}{4\pi}
\oint_\Gamma\oint_{C_{\rm mag}}
\frac{
(\bs x_1-\bs x_2)\wedge\dd\bs x_1\wedge\dd\bs x_2
}{
I\,|\bs x_1-\bs x_2|^3
}
= \pm 1.
\label{eq:gauss-link-wedge}
\end{equation}
The sign depends on the chosen orientations. To verify the linking number, deform the difference
contour, without crossing the magnetic core, to a
meridional loop around one cross-section of the
toroidal tube. A spanning disk of this meridian
intersects the oriented core curve transversely once.
Hence the oriented intersection number, and therefore
the Gauss linking number, is $\ell=\pm1$. The nonzero linking number means that $\Gamma$ is not contractible in the exterior of the magnetic core. Equivalently, a spanning surface for $\Gamma$ in all of $\R^3$ intersects the core with oriented intersection number $\ell$, whereas no spanning surface lying wholly in the flux-free exterior exists for this homology class.

\section{Flux, Connection, and the Geometric Stokes Theorem}
\label{sec:flux}

Let $S$ be an oriented surface, locally parametrised by coordinates $(u,v)$. Its directed differential area is the bivector
\begin{equation}
\dd S_2
\equiv
\frac{\partial\bs x}{\partial u}
\wedge
\frac{\partial\bs x}{\partial v}\,
\dd u\,\dd v.
\label{eq:area-element-GA}
\end{equation}
This definition is intrinsically geometric algebra: it uses only tangent 1-vectors, their exterior product, and scalar coordinate differentials. If an oriented unit normal $\hat{\bs n}$ and scalar area element $\dd S$ are introduced subsequently, then Eq.~\eqref{eq:area-element-GA} becomes $\dd S_2=I\hat{\bs n}\,\dd S$. 

The magnetic flux is the scalar pairing of the magnetic bivector with the directed surface bivector:
\begin{equation}
\boxed{
\Phi_B(S)
\equiv
-\int_S\graden{\Biv\,\dd S_2}{0}
}
\label{eq:flux-definition-GA}
\end{equation}
This is the primary geometric-algebra definition. In the auxiliary vector-normal notation just described, it is equivalent to the conventional expression because
\begin{align}
-\graden{\Biv\,\dd S_2}{0}
&=-\graden{(I\bs B)(I\hat{\bs n}\,\dd S)}{0}\nonumber\\
&=\graden{\bs B\hat{\bs n}}{0}\,\dd S
=\bs B\cdot\hat{\bs n}\,\dd S,
\label{eq:flux-vector-dual}
\end{align}
where the second line uses $(I\bs B)(I\hat{\bs n})=I^2\bs B\hat{\bs n}=-\bs B\hat{\bs n}$, since $I$ is central and $I^2=-1$.

Locally, the magnetic bivector is the exterior derivative of the potential 1-vector:
\begin{equation}
\Biv=\nabla\wedge \bs A.
\label{eq:field-potential}
\end{equation}
With the present orientation convention,
\begin{equation}
\nabla\wedge \bs A=I(\nabla\times \bs  A),
\label{eq:curl-duality}
\end{equation}
so Eq.~\eqref{eq:field-potential} is precisely the dual geometric-algebra form of the familiar relation $\bs B=\nabla\times \bs  A$. In the accessible exterior of an ideally shielded
toroid, $\Biv=0$. Consequently, $A$ is locally a
gradient: on each simply connected patch one can
write $\bs A=\nabla\chi$. The important local--global
distinction concerns this gradient representation,
not the field--potential identity itself. Because
the exterior is multiply connected, a single-valued
scalar $\chi$ need not exist on the whole exterior
region.

For an oriented surface $S$ with boundary $\partial S=\Gamma$, the
geometric Stokes theorem---obtained as the two-dimensional scalar
projection of the fundamental theorem of geometric
calculus~\cite{Hestenes1984,Macdonald2012}---gives
\begin{equation}
\oint_\Gamma \bs A\cdot\dd\bs x
=-\int_S\graden{(\nabla\wedge \bs A)\,\dd S_2}{0}.
\label{eq:stokes-GA}
\end{equation}
Substitution of Eq.~\eqref{eq:field-potential} into Eq.~\eqref{eq:stokes-GA}, followed by Eq.~\eqref{eq:flux-definition-GA}, gives
\begin{equation}
\boxed{
\oint_\Gamma \bs A\cdot\dd\bs x=\Phi_B(S)
}.
\label{eq:circulation-flux}
\end{equation}
The sign in Eq.~\eqref{eq:circulation-flux} is fixed by the common orientation convention for $\Gamma$ and $\dd S_2$. For the linked toroidal contour, the flux-free exterior alone contains no spanning surface for $\Gamma$; any spanning surface in $\R^3$ has the requisite core intersection and receives its flux contribution there.

\paragraph{Orientation check.}
The sign chain of Eqs.~\eqref{eq:stokes-GA}--\eqref{eq:circulation-flux} can be verified on a single explicit example. Let $S$ be the unit disk in the $\bs\sigma_1\bs\sigma_2$-plane, parametrised by $(u,v)$, so that $\dd S_2=\bs\sigma_1\bs\sigma_2\,\dd u\,\dd v$ and the boundary $\Gamma$ is the unit circle traversed counterclockwise. Take $A=\tfrac12(-v\,\bs\sigma_1+u\,\bs\sigma_2)$, so that $\nabla\times \bs  A=\bs\sigma_3$ and $\Biv=\nabla\wedge \bs A=I\bs\sigma_3=\bs\sigma_1\bs\sigma_2$. Then, on the one hand,
\[
\oint_\Gamma \bs A\cdot\dd\bs x=\tfrac12\oint(-v\,\dd u+u\,\dd v)=\pi,
\]
the area of the disk. On the other hand,
\[
\graden{\Biv\,\dd S_2}{0}
=\graden{\bs\sigma_1\bs\sigma_2\,\bs\sigma_1\bs\sigma_2}{0}\dd u\,\dd v
=-\dd u\,\dd v,
\]
so $-\int_S\graden{\Biv\,\dd S_2}{0}=\pi=\int_S\bs B\cdot\hat{\bs n}\,\dd S$. Both Eqs.~\eqref{eq:stokes-GA} and \eqref{eq:circulation-flux} are thereby confirmed with the stated conventions.

\section{The Central Pseudoscalar Phase and $U(1)$ Gauge Covariance}
\label{sec:phase}

The standard electromagnetic gauge group is $U(1)$. In complex notation its element is $e^{i\alpha}$, with real scalar gauge parameter $\alpha$. In $\Cl(3,0)$, the central unit pseudoscalar $I$ satisfies $I^2=-1$ and commutes with all multivectors. We represent the $U(1)$ element by
\begin{equation}
U(\alpha)
:=
\exp(I\alpha)
=
\cos\alpha+I\sin\alpha,
\qquad
U\rev U=1.
\label{eq:pseudoscalar-phase-element}
\end{equation}
In this paper, a \emph{central phase element} is an
element $U$ of the central subalgebra
$\operatorname{span}_{\R}\{1,I\}$ satisfying
$U\rev U=1$.

Such an element should not be confused with a
spatial rotor
\[
R=\exp(J\theta/2),
\qquad J^2=-1,
\]
where $J$ is a unit bivector.  A spatial rotor acts
on a vector by
\[
\bs v\mapsto R\bs v\rev R,
\]
rotating its component in the plane represented
by $J$ while leaving its perpendicular component
unchanged. The sign of the rotation angle follows
from the chosen generator convention.

The central phase element $U(\alpha)$ instead
acts on the wave amplitude by simple multiplication.
It commutes with every multivector and does not
represent a spatial rotation. Indeed,
\[
U(\alpha)\bs v\rev{U(\alpha)}=\bs v.
\]

Since $I$ is central and $\rev I=-I$, we have
\[
\rev{U(\alpha)}=U(\alpha)^{-1}=e^{-I\alpha}.
\]
The group law
\begin{equation}
U(\alpha)U(\beta)=U(\alpha+\beta)
\label{eq:U1-group-law}
\end{equation}
holds automatically.

Equation~\eqref{eq:pseudoscalar-phase-element} is therefore a faithful real-geometric-algebra representation of the Abelian gauge group $U(1)$. Up to the orientation choice $I\mapsto-I$, $I$ is the unique central element of $\Cl(3,0)$ whose square is $-1$ and which therefore generates the canonical embedded copy of $U(1)$. This represents the electromagnetic gauge group;
it does not identify gauge transformations with
physical rotations in three-dimensional space.

A spatial bivector generator is not central: it anticommutes
with vectors in its plane and commutes with vectors
perpendicular to that plane. We therefore use the central
pseudoscalar for the simple multiplicative phase convention
adopted here. This is a representation choice, not a prohibition
on right-acting bivector representations of electromagnetic
phase in other geometric-algebra formulations.
The AB phase itself does not require a physical rotation
of the electron's spin.

\subsection{Local Gauge Equivalence and Covariance}

We consider a static magnetic configuration, with no
electric field in the accessible region, and restrict
attention to time-independent gauge transformations.
The scalar potential is taken to be a spatially
constant background whose common phase contribution
is suppressed. The discussion below therefore concerns
the spatial electromagnetic connection.

Let $\mathcal M_{\rm ext}$ be the exterior region accessible to the electron. In the ideal shielding limit,
\begin{equation}
\Biv=\nabla\wedge \bs A=0
\qquad\text{on }\mathcal M_{\rm ext}.
\label{eq:exterior-flatness}
\end{equation}
On each simply connected patch $\mathcal U_a\subset\mathcal M_{\rm ext}$, there is a real scalar $\chi_a$ such that
\begin{equation}
\bs A=\nabla\chi_a.
\label{eq:local-pure-gauge}
\end{equation}
On any overlap $ \mathcal U_a\cap \mathcal U_b$ we have $\nabla(\chi_a-\chi_b)=\bs A- \bs A=0$, so $\chi_a-\chi_b$ is constant on each connected
component of the overlap. The local descriptions therefore differ by constant
central phase factors on connected overlaps. They
define consistent local descriptions and pathwise
transport, but need not combine into a globally
single-valued phase field that removes the connection.

For a wave multivector $\psi$ (an \emph{arbitrary} multivector field; no restriction to a scalar or even subalgebra is needed), the local $U(1)$ gauge transformation is fixed to be
\begin{equation}
\psi\mapsto\psi\,E_\chi,
\qquad
E_\chi\equiv\exp\!\left(I\frac{q\chi}{\hbar}\right),
\qquad
\bs A\mapsto \bs  A+\nabla\chi.
\label{eq:GA-gauge-transformation}
\end{equation}
The \emph{sign} in Eq.~\eqref{eq:GA-gauge-transformation} is dictated by the covariant derivative
\begin{equation}
D_{\bs A}\psi
\equiv
\nabla\psi-\frac{q}{\hbar}\bs A\psi I,
\label{eq:GA-covariant-derivative}
\end{equation}
and matches the standard complex transformation $\psi\mapsto e^{iq\chi/\hbar}\psi$ of the Hamiltonian $(\bs p-q\bs A)^2/2m$; with the opposite sign the $\nabla\chi$ terms below would add rather than cancel.

\begin{proposition}[Gauge covariance]
\label{prop:covariance}
For every multivector field $\psi$ and real scalar field $\chi$,
\begin{equation}
D_{\bs A+\nabla\chi}\bigl(\psi E_\chi\bigr)
=
(D_{\bs A}\psi)\,E_\chi.
\label{eq:GA-covariance}
\end{equation}
\end{proposition}

\begin{proof}
Since $E_\chi$ is central, the componentwise product rule gives
$\partial_k(\psi E_\chi)=(\partial_k\psi)E_\chi+\psi\,\partial_kE_\chi$,
with $\partial_kE_\chi=(Iq/\hbar)(\partial_k\chi)E_\chi$.
Here $\chi$ is a real scalar field, so each $\partial_k\chi$ is scalar,
while $I$ and $E_\chi$ are central.
Assembling the vector derivative $\nabla=\sum_k\bs\sigma_k\partial_k$ gives
\begin{align}
\nabla(\psi E_\chi)
&=\sum_k\bs\sigma_k\,\partial_k(\psi E_\chi)\nonumber\\
&=(\nabla\psi)E_\chi+\frac{Iq}{\hbar}\sum_k\bs\sigma_k\,\psi\,(\partial_k\chi)\,E_\chi.
\label{eq:product-rule-central}
\end{align}
The scalar coefficients may be moved through the geometric product, so
\begin{equation}
\sum_k\bs\sigma_k\,\psi\,(\partial_k\chi)
=\sum_k(\partial_k\chi)\,\bs\sigma_k\psi
=(\nabla\chi)\psi.
\label{eq:gradient-remains-left}
\end{equation}
No commutation of $\nabla\chi$ with $\psi$ is required: $\nabla\chi$ remains on the left in both expressions.
Centrality of $I$ and $E_\chi$ then gives
\begin{equation}
\nabla(\psi E_\chi)
=(\nabla\psi)E_\chi+\frac{q}{\hbar}(\nabla\chi)\psi E_\chi I.
\label{eq:product-rule-phase}
\end{equation}
Therefore
\begin{align}
D_{\bs A+\nabla\chi}(\psi E_\chi)
&=(\nabla\psi)E_\chi+\frac{q}{\hbar}(\nabla\chi)\psi E_\chi I
-\frac{q}{\hbar}(A+\nabla\chi)\psi E_\chi I\nonumber\\
&=(\nabla\psi)E_\chi-\frac{q}{\hbar} \bs A\psi E_\chi I\nonumber\\
&=(D_{\bs A}\psi)E_\chi.
\label{eq:GA-covariance-proof}
\end{align}
The two $\nabla\chi$ terms cancel.
The cancellation above uses the assumption that $\chi$ is
real scalar-valued. A general multivector-valued parameter
does not preserve this argument or, in general, the vector
character of the transformed potential. We therefore
restrict throughout to the standard real scalar
electromagnetic gauge parameter.
\end{proof}

Assume that the quantum evolution equation is
gauge covariant with the stated minimal-coupling
convention. A local solution $\psi_0$ in the gauge
$\bs A=0$ on $\mathcal U_a$ is then mapped to a solution
in the gauge $\bs A=\nabla\chi_a$ by
Eq.~\eqref{eq:GA-gauge-transformation}, giving
\begin{equation}
\psi_a=\psi_0U_a,
\qquad
U_a(\bs x)
=
\exp\!\left(I\frac{q\chi_a(\bs x)}{\hbar}\right).
\label{eq:local-phase-rotor}
\end{equation}
Differentiation of Eq.~\eqref{eq:local-phase-rotor} gives the transport equation
\begin{equation}
\nabla U_a
=
\frac{q}{\hbar} \bs AU_aI
\qquad\text{on } \mathcal U_a,
\label{eq:local-transport-equation}
\end{equation}
since $\nabla E_\chi=(Iq/\hbar)(\nabla\chi)E_\chi=(q/\hbar)(\nabla\chi)E_\chi I$ and $\bs A=\nabla\chi_a$ on the patch. Applying Proposition~\ref{prop:covariance} to the
transformation from $\bs A=0$ to $\bs A=\nabla\chi_a$ gives
\begin{equation}
D_{\bs A}(\psi_0U_a)=(\nabla\psi_0)U_a.
\label{eq:local-free-equivalence}
\end{equation}

Equation~\eqref{eq:local-free-equivalence} expresses
local gauge equivalence in the field-free exterior.
A globally single-valued phase can remove the
connection only when its holonomy is trivial
around every closed loop. Multiply connected
geometry permits, but does not require,
nontrivial holonomy.

The covariance proposition applies algebraically
to arbitrary multivector fields; it does not
identify them all with physical electron states.
For interference we use only
$\psi=zR$, where
$z\in\operatorname{span}_{\R}\{1,I\}$ and the
same even multivector $R$, with $R\rev R=1$,
carries the common spin-frame data. Then
\[
\langle\psi\rev\psi\rangle_0
=z\rev z=|z|^2.
\]
This embeds the usual common-spin scalar-amplitude
calculation; no general spin dynamics is derived.

\subsection{Holonomy}

Let $\gamma$ run from $\bs x_i$ to $\bs x_f$, contained in a single
simply connected patch $\mathcal U_a$ with $\bs A=\nabla\chi_a$. Recall the local
transport equation~\eqref{eq:local-transport-equation},
\begin{equation}
\nabla U_a=\frac{q}{\hbar} \bs AU_aI,
\label{eq:transport-recall}
\end{equation}
understood componentwise as $\partial_kU_a=\frac{q}{\hbar}A_kU_aI$ for
each spatial direction $\bs\sigma_k$. Parametrise $\gamma$ by arc
length or any convenient parameter $s$, with tangent $\dd\bs x/\dd s$.
The chain rule gives the rate of change of $U_a$ along $\gamma$:
\begin{align}
\frac{\dd}{\dd s}U_a(\gamma(s))
&=\sum_k\frac{\dd x_k}{\dd s}\,\partial_kU_a
\nonumber\\
&=\frac{q}{\hbar}
\left(\bs A\cdot\frac{\dd\bs x}{\dd s}\right)U_aI.
\label{eq:chain-rule-path}
\end{align}

Multiplying through by $\dd s$ gives the differential path
equation
\begin{equation}
\dd U_a=\frac{q}{\hbar}(\bs A\cdot\dd\bs x)\,U_aI.
\label{eq:path-differential}
\end{equation}
On a path $\gamma$ contained in $\mathcal U_a$, let
$s_i$ be the initial parameter value, with
$\gamma(s_i)=\bs x_i$. The local phase element
$U_a(\bs x)$ need not equal $1$ at $\bs x_i$.
To describe only the phase accumulated from the
starting point, define
\[
U_\gamma(s)
=
U_a(\gamma(s))\,U_a(\bs x_i)^{-1}.
\]
At the start of the path,
\[
U_\gamma(s_i)
=
U_a(\bs x_i)\,U_a(\bs x_i)^{-1}
=1.
\]
Thus the initial value $1$ is a choice of phase
reference, not a statement that the potential or
the local phase element vanishes or is trivial.

The starting point $\bs x_i$ is fixed as $s$ varies.
Hence
\[
C_i:=U_a(\bs x_i)^{-1}
\]
is a constant, even though $U_a(\gamma(s))$ generally
changes along the path. Writing
$U_\gamma(s)=U_a(\gamma(s))C_i$ and applying the
product rule gives
\begin{align*}
\frac{\dd U_\gamma}{\dd s}
&=
\left[
\frac{\dd}{\dd s}U_a(\gamma(s))
\right]C_i\\
&=
\frac{q}{\hbar}
\left(
\bs A(\gamma(s))\cdot\frac{\dd\gamma}{\dd s}
\right)
U_a(\gamma(s))I C_i\\
&=
\frac{q}{\hbar}
\left(
\bs A(\gamma(s))\cdot\frac{\dd\gamma}{\dd s}
\right)
U_a(\gamma(s))C_i I\\
&=
\frac{q}{\hbar}
\left(
\bs A(\gamma(s))\cdot\frac{\dd\gamma}{\dd s}
\right)
U_\gamma(s)I.
\end{align*}
Here $I C_i=C_i I$, since $I$ is central.
Thus removing the fixed starting phase changes
the initial value to $U_\gamma(s_i)=1$ without
changing the transport equation. The value $1$
means that no phase has yet accumulated relative
to the starting point; it does not require the
subsequent rate of phase accumulation to vanish.

For a path crossing several patches, solve this
equation successively on each segment, using the
final value on one segment as the initial value
on the next. The phase is not reset to $1$ at
patch boundaries. Consequently,
\begin{equation}
\dd U_\gamma
=
\frac{q}{\hbar}(\bs A\cdot\dd\bs x)\,U_\gamma I,
\qquad
U_\gamma(s_i)=1.
\label{eq:U-path-equation}
\end{equation}

\paragraph{Consistency check.}
Because the gauge group is Abelian and $I$ is central,
no path ordering is required. At an intermediate
parameter value $s$, the solution is
\[
U_\gamma(s)
=
\exp\!\left[
I\frac{q}{\hbar}
\int_{s_i}^{s}
A(\gamma(t))\cdot\frac{\dd\gamma}{\dd t}\,\dd t
\right].
\]
The integral is zero at $s=s_i$, so
$U_\gamma(s_i)=1$. Differentiation with respect
to the upper limit gives
\[
\frac{\dd U_\gamma}{\dd s}
=
I\frac{q}{\hbar}
\left(
\bs A(\gamma(s))\cdot\frac{\dd\gamma}{\dd s}
\right)U_\gamma(s),
\]
which is the stated transport equation, since
$I$ commutes with $U_\gamma$.

At the final parameter value $s_f$, we abbreviate
the complete transport phase by
$U_\gamma:=U_\gamma(s_f)$. Thus
\begin{equation}
U_\gamma
=
\exp\!\left(
I\frac{q}{\hbar}\int_\gamma \bs A\cdot\dd\bs x
\right).
\label{eq:path-rotor-check}
\end{equation}

Under the gauge transformation~\eqref{eq:GA-gauge-transformation}, $\chi_a\mapsto\chi_a+\chi$, hence $U_a\mapsto U_aE_\chi$ and
\begin{equation}
U_\gamma\;\mapsto\;
E_\chi(\bs x_f)\,U_\gamma\,E_\chi(\bs x_i)^{-1}.
\label{eq:path-gauge-transform}
\end{equation}
Thus an open-path transport phase depends on the
gauge through its endpoint factors. For two paths
with common endpoints, these factors cancel in
their relative transport phase, giving the
gauge-invariant closed-loop holonomy.

Define the relative transport, or closed-loop holonomy, by
\[
U_\Gamma:=U_{\gamma_1}\rev U_{\gamma_2}.
\]
Using the group law~\eqref{eq:U1-group-law},
the circulation--flux identity~\eqref{eq:circulation-flux},
and $\Gamma=\gamma_1-\gamma_2$,
\begin{align}
U_\Gamma
&=U_{\gamma_1}\rev U_{\gamma_2}\nonumber\\
&=
\exp\!\left(
I\frac{q}{\hbar}
\oint_{\gamma_1-\gamma_2}\bs A\cdot\dd\bs x
\right)\nonumber\\
&=
\boxed{
\exp\!\left(I\frac{q\Phi_B}{\hbar}\right)
}.
\label{eq:holonomy}
\end{align}
The gauge-invariant content is the closed-loop holonomy element, or equivalently its angle $q\Phi_B/\hbar$ modulo $2\pi$. This is consistent with and fixes the sign convention used in Eq.~\eqref{eq:AB-formula}.

\paragraph{Remark (comparison with spacetime algebra).}
In the spacetime algebra formulation of Dirac theory,\footnote{The spacetime algebra formulation uses $\Cl(1,3)$ rather than $\Cl(3,0)$ as its base Clifford algebra.}
electromagnetic phase can be represented by a right-acting
bivector generator, conventionally written $I\bs\sigma_3$
\cite{Hestenes1966,DoranLasenby2003}.
The symbol $I$ there denotes the spacetime pseudoscalar,
and the $\bs\sigma_k$ are relative spatial vectors.
These objects should not be identified directly with the
grade assignments used in the present $\Cl(3,0)$ construction.
Here the central pseudoscalar gives a particularly simple
realization of complex phase, while the componentwise
calculation above establishes gauge covariance.

\section{Interference Pattern and Its Justification}
\label{sec:interference}

Given the usual Born rule and the common-spin assumptions
below, the interference formula follows directly from the scalar part of the detector density multivector. We assume that the two arriving electron components have
the same local spin polarization and coherent spatial
overlap at the detector, differing only in scalar
amplitude and relative \(U(1)\) phase.  Path-dependent
spin rotations and imperfect spatial overlap are not
modelled here.  Conditional source--field records and
their effect on coherence are considered in
Section~\ref{sec:backreaction}.

Let $a_1,a_2\geq0$ be the branch amplitudes, let $R$ be the common normalized even multivector encoding the shared local spin state and frame ($R\rev R=1$), and let $\delta_0$ be the ordinary, non-topological relative propagation phase. Since $a_1,a_2$ are real scalars and $I$ is central, the branch contributions may be written
\begin{equation}
\psi_1=a_1R,
\qquad
\psi_2=a_2Re^{I\delta_0}=a_2e^{I\delta_0}R.
\label{eq:branch-amplitudes}
\end{equation}
The two physical branch contributions carry their respective
transport phases:
\[
\psi_{\rm det}^{\rm phys}
=\psi_1U_{\gamma_1}+\psi_2U_{\gamma_2}.
\]
Because these phase elements are central, and
$U_\Gamma=U_{\gamma_1}\rev U_{\gamma_2}$, recalling that $\rev U_{\gamma_2}=U_{\gamma_2}^{-1}$, this becomes
\[
\psi_{\rm det}^{\rm phys}
=\left(\psi_1+\psi_2U_\Gamma^{-1}\right)U_{\gamma_1}.
\]
The common unit phase $U_{\gamma_1}$ does not change the
detection probability. We may therefore use the equivalent
detector representative
\begin{equation}
\psi_{\rm det}
=\psi_1+\psi_2U_\Gamma^{-1}
=\left[a_1+a_2e^{I\Theta}\right]R,
\qquad
\Theta\equiv\delta_0-\frac{q\Phi_B}{\hbar}.
\label{eq:detector-state}
\end{equation}
Here $\delta_0$ is the ordinary phase of branch 2 relative
to branch 1, whereas $q\Phi_B/\hbar$ is the electromagnetic
phase of branch 1 relative to branch 2.

Using $R\rev R=1$ and centrality of $I$, together with
$\rev{(e^{I\Theta})}=e^{-I\Theta}$ and the antihomomorphism property of
reversion, the geometric-algebra form of the Born rule gives the
detection probability density. For an effective detector-plane amplitude normalized
with respect to $\dd A$, $\mathcal P(\bs x)\,\dd A$
is the probability of detection in the area element
$\dd A$. For an ensemble of identically prepared electrons, the
expected counts are proportional to this probability.

\begin{align}
\mathcal P:=\graden{\psi_{\rm det}\rev{\psi_{\rm det}}}{0}
&=
\graden{
\left[a_1+a_2e^{I\Theta}\right]R
\rev R
\left[a_1+a_2e^{-I\Theta}\right]
}{0}\nonumber\\
&=
\graden{
a_1^2+a_2^2+a_1a_2
\left(e^{I\Theta}+e^{-I\Theta}\right)
}{0}\nonumber\\
&=
a_1^2+a_2^2+2a_1a_2\cos\Theta,
\label{eq:interference-derived}
\end{align}
since $e^{I\Theta}+e^{-I\Theta}=2\cos\Theta$ is already scalar. Equation~\eqref{eq:interference-derived} is the standard two-beam interference law in geometric-algebra form. In the equal-amplitude case $a_1=a_2=\sqrt{\rho_0}$, the detector intensity is

\begin{equation}
\mathcal P
=2\rho_0
\left[
1+\cos\left(\delta_0-\frac{q\Phi_B}{\hbar}\right)
\right].
\label{eq:interference-equal-branches}
\end{equation}
The AB contribution therefore changes the relative phase
appearing in the interference term by $-q\Phi_B/\hbar$
with the branch convention adopted here.
Reversing the orientation of $\Gamma$ reverses the sign
of the quoted AB phase.

\section{Superconducting Shield and Binary Phase Parity}
\label{sec:superconductivity}

Let the superconducting condensate have Cooper-pair charge $q_c=-2e$ and effective pair mass $m^*$. Consistently with Eq.~\eqref{eq:GA-gauge-transformation}, its order parameter transforms as $\Psi_{\rm sc}\mapsto\Psi_{\rm sc}\exp(Iq_c\chi/\hbar)$ and may be written
\begin{equation}
\Psi_{\rm sc}(\bs x)
=
f(\bs x)\,e^{I\theta_{\rm sc}(\bs x)},
\qquad
f:=\sqrt{n_s}\geq0.
\label{eq:GL-order}
\end{equation}

\paragraph{Physical meaning of the order parameter.}
In a conventional superconductor, electrons form
Cooper pairs of charge $q_c=-2e$. The superconducting
state has a collective amplitude and phase, described
by the macroscopic order parameter $\Psi_{\rm sc}$
\cite{Tinkham1996}.
Here we choose its normalization so that
\[
|\Psi_{\rm sc}|^2
\equiv
\Psi_{\rm sc}\widetilde{\Psi_{\rm sc}}
=n_s,
\]
where $n_s$ is the effective number of superconducting
pairs per unit volume. Thus $f=\sqrt{n_s}$ is simply
the amplitude whose square gives this density.
The factor $e^{I\theta_{\rm sc}}$ describes the
collective phase. This is a field describing the
condensate, not a single-electron probability
amplitude.

\paragraph{Where the supercurrent comes from.}
Electric current is moving charge. If the pairs
have local mean velocity $\bs v_s$, their
charge-current density is
\[
\bs j_s=q_c n_s\bs v_s.
\]
This is the usual relation ``charge per carrier
times number density times velocity''. Because
$q_c<0$, the conventional current points opposite
to the motion of the pairs.

To connect this flow with the condensate phase,
we use the standard effective-mass description
of superconductivity \cite{Tinkham1996}:
\[
m^*\bs v_s
=
\hbar\nabla\theta_{\rm sc}-q_c \bs A.
\]
The right-hand side is the gauge-invariant
mechanical momentum of a pair. A phase gradient
can therefore describe a condensate flow, but
in an electromagnetic field the phase gradient
and vector potential must be considered together.
Substituting this velocity into the current
formula gives
\[
\bs j_s
=
\frac{q_c n_s}{m^*}
\left(\hbar\nabla\theta_{\rm sc}-q_c \bs A\right).
\]
We take this standard condensate relation as a
physical input; the geometric-algebra calculation
below expresses it in the notation of this paper.

The gauge-covariant derivative, from Eq.~\eqref{eq:GA-covariant-derivative} with $q\to q_c$, is
\begin{equation}
D_{\bs A}\Psi_{\rm sc}
=
\nabla\Psi_{\rm sc}
-\frac{q_c}{\hbar}\bs A\Psi_{\rm sc}I
=
\nabla\Psi_{\rm sc}
+\frac{2e}{\hbar}\bs A\Psi_{\rm sc}I.
\label{eq:GL-covariant}
\end{equation}

Expanding with the componentwise product rule ($f$ is a scalar field, $I$ central),
\begin{align}
\nabla\Psi_{\rm sc}
&=\sum_k\bs\sigma_k\,\partial_k\!\left(fe^{I\theta_{\rm sc}}\right)
=\bigl[\nabla f+If\,\nabla\theta_{\rm sc}\bigr]e^{I\theta_{\rm sc}},
\end{align}
so that
\begin{align}
D_{\bs A}\Psi_{\rm sc}
&=\left[
\nabla f+If\,\nabla\theta_{\rm sc}
-\frac{Iq_c}{\hbar}fA
\right]e^{I\theta_{\rm sc}}
\nonumber\\
&=\left[
\nabla f+\frac{If}{\hbar}\bs p_s
\right]e^{I\theta_{\rm sc}},
\label{eq:GL-expansion}
\end{align}
where the \emph{gauge-invariant superfluid momentum} 1-vector is
\begin{equation}
\bs p_s
\equiv
\hbar\nabla\theta_{\rm sc}-q_c \bs A
=
\hbar\nabla\theta_{\rm sc}+2e \bs A.
\label{eq:superfluid-momentum}
\end{equation}

Under a gauge transformation,
$\theta_{\rm sc}\mapsto\theta_{\rm sc}+q_c\chi/\hbar$
and $\bs A\mapsto \bs A+\nabla\chi$, so the two changes
in $\bs p_s$ cancel.

The \emph{charge} supercurrent density is
\begin{equation}
\bs j_s
\equiv
q_c n_s\bs v_s
=
\frac{q_c n_s}{m^*}\bs p_s
=
-\frac{2e\,n_s}{m^*}\bs p_s,
\qquad
\bs v_s=\frac{\bs p_s}{m^*}.
\label{eq:supercurrent}
\end{equation}

For a closed contour $C$ inside the superconducting material and encircling the magnetic core once, single-valuedness requires
\begin{equation}
\oint_C\nabla\theta_{\rm sc}\cdot\dd\bs x
=2\pi n_{\rm w},
\qquad n_{\rm w}\in\Z
\label{eq:phase-winding}
\end{equation}
Here $C$ is a meridional contour around the magnetic
tube cross-section. We choose its orientation to agree
with the linking orientation of the exterior electron
contour $\Gamma$. In the ideal shielding limit, we
assume that these contours bound an annular surface
$\Sigma$ lying in the field-free superconducting bulk
and exterior region, with oriented boundary
$\partial\Sigma=\Gamma-C$. Stokes' theorem then gives
\[
\oint_\Gamma \bs A\cdot\dd\bs x
-\oint_C \bs A\cdot\dd\bs x
=
-\int_\Sigma\graden{\Biv\,\dd S_2}{0}
=0.
\]
Thus the two contours measure the same linked magnetic
flux in this idealization. Finite penetration or leakage
fields can produce a nonzero flux through $\Sigma$,
and hence a difference between the two circulations.

We choose $C$ within the superconducting material where
$n_s>0$ everywhere along the contour, so that the
condensate phase and its winding are well defined. Combining Eqs.~\eqref{eq:superfluid-momentum} and \eqref{eq:phase-winding} step by step:
\begin{align}
2\pi n_{\rm w}\hbar
&=\oint_C\hbar\nabla\theta_{\rm sc}\cdot\dd\bs x
=\oint_C\bigl(\bs p_s-2e\bs A\bigr)\cdot\dd\bs x\nonumber\\
&=-\frac{m^*}{2e}\oint_C\ (\bs j_s/n_s)\cdot\dd\bs x
\;-\;2e\,\Phi_B(C),
\label{eq:fluxoid-derivation}
\end{align}
where Eq.~\eqref{eq:supercurrent} was used to eliminate $\bs p_s$, and
$\Phi_B(C)\equiv\oint_C\bs A\cdot\dd\bs x$ is the flux through any spanning surface, by Eq.~\eqref{eq:circulation-flux}. Dividing Eq.~\eqref{eq:fluxoid-derivation} by $-2e$ gives the fluxoid quantization condition

\begin{equation}
\boxed{
\oint_C
\left(
\bs A+\mu_0\lambda_L^2\bs j_s
\right)\cdot\dd\bs x
=N\Phi_0,
\qquad
N:=-n_{\rm w},
\qquad
\Phi_0:=\frac{h}{2e}.
}
\label{eq:fluxoid}
\end{equation}

The London penetration depth $\lambda_L$ is the
characteristic distance over which a weak magnetic
field is screened inside the superconductor.
In the effective-mass description used here,
\cite{Tinkham1996}
\begin{equation}
\mu_0\lambda_L^2
=\frac{m^*}{n_s q_c^2}
=\frac{m^*}{4e^2n_s}.
\label{eq:london}
\end{equation}

Single-valuedness of the nonvanishing condensate order
parameter gives an exact integer phase winding.
Within the condensate momentum--current description
used here, this yields the fluxoid quantization
condition above. When the superconducting shield is sufficiently thick
compared with the relevant magnetic penetration depth,
one may choose a contour $C$ well inside the material,
away from screening layers, on which the current
contribution is negligible. The fluxoid condition
then reduces to the approximation
\begin{equation}
\Phi_B(C)\simeq N\Phi_0,
\qquad
\Phi_0=\frac{h}{2e}.
\label{eq:flux-quantization}
\end{equation}

The field-free annulus is an additional ideal-limit
assumption. For a finite penetration depth, a connecting
surface may cross a screening layer, so equality of the
two circulations must be checked rather than inferred
solely from their equal linking number. When the flux through the connecting annulus also
vanishes, this is the flux linked by the exterior
electron contour $\Gamma$.

Fluxoid quantization does not, in general, imply exact
quantization of the magnetic flux. The exactly binary
holonomy below refers to the ideal limit
$\Phi_B=N\Phi_0$, where the screening-current correction
vanishes. For finite corrections, the phase is
correspondingly only approximately parity-valued.

\subsection{Single-Electron Binary Contrast}

For an electron probe, $q=-e$. We now impose the
exact ideal limit $\Phi_B=N\Phi_0$, with the
screening-current contribution and the flux difference
between $C$ and $\Gamma$ both vanishing. Then
\begin{align}
\frac{q\Phi_B}{\hbar}
&=
\frac{-e}{\hbar}
\left(N\frac{h}{2e}\right)
\nonumber\\
&=-N\pi.
\label{eq:electron-fluxoid-phase}
\end{align}
The electron holonomy element in Eq.~\eqref{eq:holonomy} becomes
\begin{equation}
U_\Gamma
=
\exp(-IN\pi)
=
\cos(N\pi)-I\sin(N\pi)
=(-1)^N.
\label{eq:parity}
\end{equation}
For a finite departure
$\Phi_B=N\Phi_0+\delta\Phi$, the corresponding
holonomy is instead
\[
U_\Gamma
=(-1)^N\exp\!\left(-I\frac{e\,\delta\Phi}{\hbar}\right).
\]
The parity approximation is accurate when
$|e\,\delta\Phi/\hbar|\ll1$.
The ideal even--odd phase contrast is the
characteristic superconducting-toroid result
\cite{Tonomura1986,Osakabe1986}.

\paragraph{Physical interpretation.}
The superconducting shield and the electron probe
respond to the same linked flux, but carry different
charges. Single-valuedness of the Cooper-pair order
parameter constrains its phase winding to an integer
multiple of $2\pi$, producing fluxoid quantization
with flux scale $\Phi_0=h/(2e)$. This does not imply
that the current vanishes everywhere: screening
currents can remain near the shield surfaces.

If the selected contour lies sufficiently deep in
the superconducting bulk that its current correction
is negligible, and if it links the same flux as the
electron contour, the linked magnetic flux approaches
$N\Phi_0$. An electron of charge $-e$ then acquires
the phase
\[
\frac{-eN\Phi_0}{\hbar}=-N\pi.
\]
Thus each superconducting flux quantum contributes
a factor $-1$ to the electron holonomy. Even $N$
gives unit holonomy, whereas odd $N$ reverses the
relative sign of the two electron amplitudes,
shifting the interference pattern by half a fringe
period. This is a relative path phase, not a
rotation of the electron's spin.

\section{Quantum Source and Backreaction}
\label{sec:backreaction}

The preceding treatment describes a quantum electron in a
classical electromagnetic background. A more complete
description also treats the magnetic source and field
quantum mechanically. Such treatments recover the usual
AB phase in the appropriate background limit, while allowing
the source and field to respond to the passing electron
\cite{Vaidman2012,PearleRizzi2017A,PearleRizzi2017B,MarlettoVedral2020}.

This section parametrises the effect of conditional
source--field states; it does not calculate their
overlap from a microscopic dynamical model. We use standard Dirac notation to label quantum states, but
represent their scalar phase coefficients by the central
Clifford subalgebra $\operatorname{span}_{\R}\{1,I\}$.
Thus $e^{I\alpha}$ acts as a scalar phase on a ket, and
reversion of these coefficients gives their conjugates.
This notation does not restrict the source and field
to a finite-dimensional Clifford algebra.

Let $|S_1\rangle$ and $|S_2\rangle$ be normalized final
source--field states conditional on the two electron paths.
At a detector position, after suppressing the common
arriving electron spin state, the combined amplitude is
\[
|F\rangle
=a_1|S_1\rangle
+a_2e^{I\Theta}|S_2\rangle,
\qquad
\Theta=\delta_0-\frac{q\Phi_B}{\hbar}.
\]
The amplitudes $a_1,a_2$ include the splitting and propagation
factors. The background AB phase has been displayed explicitly
and is not included again in the residual source response.

Represent the conditional-state overlap by
\[
\Omega_s:=\langle S_1|S_2\rangle
=\nu e^{I\delta_s},
\qquad 0\leq\nu\leq1.
\]
Here $\nu$ measures the overlap of the two final quantum
states, and $\delta_s$ is a possible additional relative
phase. When $\nu=0$, this phase is immaterial.

Only the combined phase $\Theta+\delta_s$ enters
the interference term; its allocation between the
branch coefficients and the conditional states
depends on their phase convention.
No nonzero residual phase shift or visibility loss
is predicted here for the Tonomura apparatus without
a dynamical calculation of $\Omega_s$.

The detector does not resolve the source and field states.
Using $\langle S_2|S_1\rangle=\rev{\Omega_s}$, its probability
density is
\begin{align}
\mathcal P_{\rm det}
&=a_1^2+a_2^2+
a_1a_2\left(
e^{I\Theta}\Omega_s+
\rev{e^{I\Theta}\Omega_s}
\right)
\nonumber\\
&=a_1^2+a_2^2+
2a_1a_2\graden{e^{I\Theta}\Omega_s}{0}
\nonumber\\
&=a_1^2+a_2^2+
2a_1a_2\nu
\cos\left(
\delta_0-\frac{q\Phi_B}{\hbar}+\delta_s
\right).
\label{eq:visibility-with-source}
\end{align}

If the alternatives leave the same final source and field
state, then $\Omega_s=1$ and the original AB interference
formula is recovered. Conditional states differing only by a phase retain
$\nu=1$. In a fixed convention for the branch
coefficients, this phase enters the interference
term; only the combined relative phase is
convention-independent.  Distinguishable
records reduce their overlap, and orthogonal conditional
states eliminate the interference term
\cite{BreuerPetruccione2002,Schlosshauer2007}.

For fixed branch amplitudes and overlap while the phase
is scanned, the visibility is
\[
V=\frac{2a_1a_2}{a_1^2+a_2^2}\,\nu,
\]
which equals $\nu$ for balanced branches.
Thus a quantum description retains the familiar AB shift
in the appropriate limit, while allowing a backreaction
to change the phase and attenuate the fringes.
Preservation of a fluxoid sector alone does not guarantee
identical final source and field states.

Here we assume a pure initial source--field state, or include
a purifying environment. 

\section{Conclusion}

The shielded toroidal AB experiment connects
confined magnetic flux with interference along
field-free electron paths. In $\Cl(3,0)$,
the magnetic bivector and directed surface
bivector encode flux geometry, while the central
pseudoscalar represents electromagnetic phase.
Their distinct roles lead to the gauge-invariant
holonomy
\[
U_\Gamma=\exp(Iq\Phi_B/\hbar).
\]

Superconducting phase winding quantises the
fluxoid. When its current correction and the
flux difference between the superconducting
and electron contours vanish, the linked flux
is $N h/(2e)$ and the electron holonomy is
$(-1)^N$. The conditional-state extension
separately shows how source--field records
can modify phase and visibility without
altering this background derivation.

\section*{Acknowledgements}

The author would like to acknowledge extremely helpful comments from Sonia Contera on an earlier draft, and Perplexity AI for help with the drafting of the paper. The author has revised the paper and remains responsible for any remaining errors.

\end{document}